\documentclass[aps,prl,showpacs,twocolumn]{revtex4-1}

\usepackage{color}
\usepackage[utf8]{inputenc}
\usepackage[english,british]{babel}
\usepackage{graphicx}
\usepackage{hyperref}
\usepackage{epstopdf}
\usepackage{amsfonts}
\usepackage{amsmath}

\newcommand{\ie}{i.\,e.\ }

\newcommand{\cf}{cf.\ }

\newcommand{\abs}[1]{|#1|}
\newcommand{\ii}{\mathrm{i}}
\newcommand{\ee}{\mathrm{e}}

\newcommand{\fref}[1]{\text{fig.}~\ref{#1}}

\begin{document}

\title{Atomic selfordering in a ring cavity with counterpropagating  pump}

\author{S. Ostermann}
\email{stefan.ostermann@uibk.ac.at}
\author{T. Grie{\ss}er}
\author{H. Ritsch}
\affiliation{Institut f\"ur Theoretische Physik, Universit\"at Innsbruck, Technikerstraße 25, A-6020 Innsbruck, Austria}

\date{\today}

\begin{abstract}
The collective dynamics of mobile scatterers and light in optical resonators generates complex behaviour. For strong transverse illumination a phase transition from homogeneous to crystalline particle order appears. In contrast, a gas inside a single-side pumped ring cavity exhibits an instability towards bunching and collective acceleration called collective atomic recoil lasing (CARL). We demonstrate that by driving two orthogonally polarized counter propagating modes of a ring resonator one realises both cases within one system. The corresponding phase diagram depending on the two pump intensities exhibits regions in which either a generalized form of self-ordering towards a travelling density wave with constant centre of mass velocity or a CARL instability is formed. Controlling the cavity driving then allows to accelerate or slow down and trap a sufficiently dense beam of linearly polarizable particles.
\end{abstract}


\maketitle

\section{Introduction}
Ultracold particles in an optical resonator interact non-locally via collective scattering of photons in and out of the cavity modes.  Under suitable conditions this induces collective instabilities\cite{bonifacio1994exponential} or even crystallisation of the particles~\cite{ritsch2013cold,baumann2010dicke,gopalakrishnan2010viewpoint}. One of the earliest examples of such an instability, developed in close analogy to free electron lasers~\cite{hopf1976classical},  was studied in the so called collective atomic recoil lasing (CARL)~\cite{bonifacio1994exponential,schmidt2014dynamical,bux2013control}. This type of instability can be realised in a single side pumped ring cavity and it reveals a transient bunching concurrent with coherent collective backscattering of pump light for an ensemble of fast particles counterpropagating the pump field of the cavity.
In an alternative geometry, considering cold particles with transverse pump in a standing wave cavity, a phase transition from homogeneous to crystalline order was predicted~\cite{domokos2002collective}  and experimentally verified~\cite{black2003observation,arnold2012self}. Later this was identified and as well confirmed as a quantum phase transition also occurring at zero temperature~\cite{nagy2010dicke,baumann2010dicke}.

In this paper we show that in a generalized geometry using two counter propagating pump fields of orthogonal polarization (CARLO), a very similar type of phase transition appears, where the system breaks its translational symmetry and transforms into periodic order. The geometry is related to the configuration studied  in~\cite{ostermann2014scattering}, where no cavity was present.  It is important to note that the pump fields injected from two sides into the ring cavity do not interfere and hence do not form a prescribed optical lattice, as they have orthogonal polarization. A lattice only appears through interference of pump and backscattered light. The two fields of orthogonal polarization interact only indirectly by scattering from the same atomic density distribution~\cite{ostermann2014scattering}.

This work is organized as follows: After a short presentation of the model, we study general properties of the system and exhibit its relation to known models. In particular using a Vlasov-type approach we study the stability boundary of the homogeneous distribution. To understand the system's behaviour in more detail, we perform specific numerical simulations in part two. We reveal that selfordered solutions with a constant centre of mass velocity can be realised. In addition we show that the system allows for slowing down a fast atomic or molecular beam. In the last part we derive expressions which enable to state whether the system settles in a selfordered phase or a  CARL instability depending on the pump parameters.
\begin{figure}
\centering
\includegraphics[width=65mm]{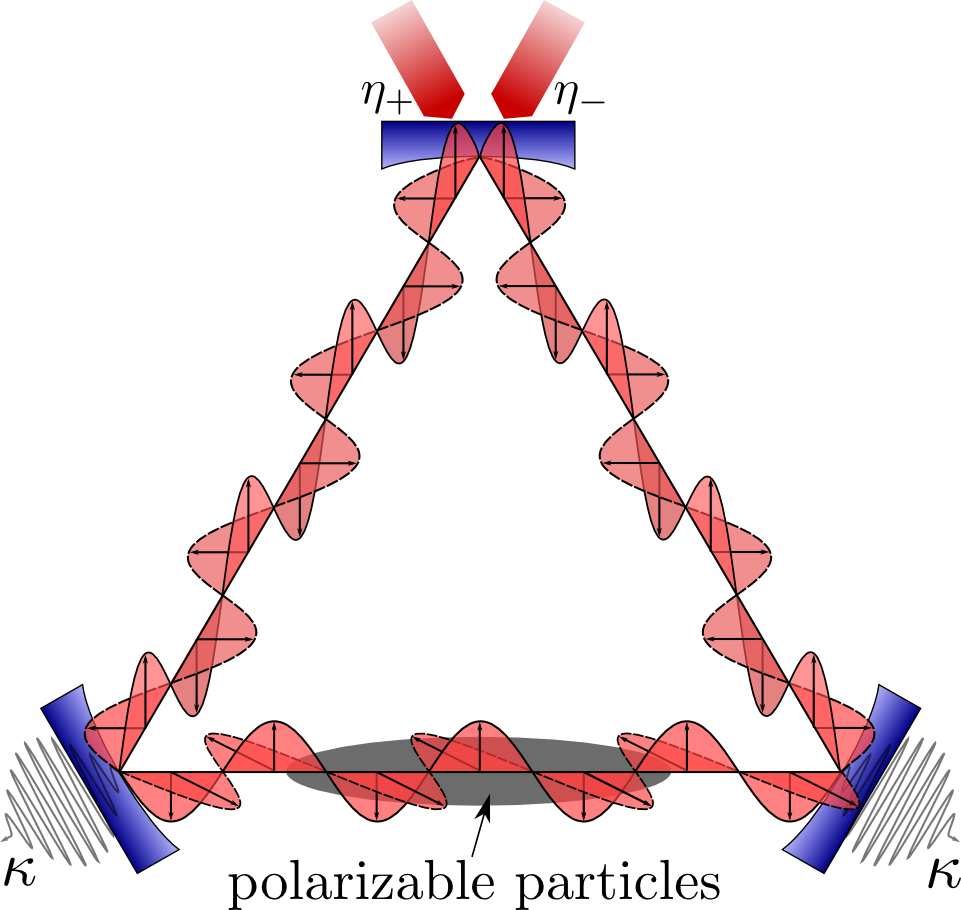}
\caption{(Colour online)  Schematic picture of the considered setup.}
\label{fig:scheme}
\end{figure}

\section{Model}
Let us consider a large ensemble of $N$ polarizable particles within a ring cavity supporting pairs of orthogonally polarized counterpropagating modes. For simplicity, we assume them to be linearly polarizable with a real scalar polarizability. As optically pumped atoms often have tensor polarizabilities, where the two orthogonal polarizations couple to different atomic transitions, the equations below have to be adapted for this case. However, for sufficiently large detuning from the optical resonances, we can largely neglect mode mixing due to spontaneous Raman transitions to other Zeeman levels and thus we simply end up with an effective polarizability for each field mode. Note that in this case we do not include optical pumping and polarization gradient cooling as in optical molasses. When enhanced by cavity feedback this would tend to localize the particles in space as investigated in some earlier work~\cite{gangl2001cavity}. While this is certainly a very interesting generalization and extension of our model, such ordering is a single-particle effect an thus fundamentally different from the collective selfordering dynamics into a lattice structure as studied below.

In a semiclassical point particle description the time evolution of the mode amplitudes $a_n$ of the intra cavity field $\mathbf{E}(\mathbf{x}):=\sum_na_n \mathbf{f}_n(\mathbf{x})$ is governed by the equations~\cite{salzburger2009collective}
\begin{equation}
\dot{a}_n(t)=(i\Delta_c-\kappa)a_n(t)
-\ii U_0\sum_{j=1}^N \mathbf{E(\mathbf{x}}_j)\!\cdot\!\mathbf{f}_n^*(\mathbf{x}_j)+\eta_n
\label{eqn:modesgeneral}
\end{equation}
where $2\kappa$ is the cavity linewidth and $\Delta_c:=\omega_p-\omega_c$ denotes the detuning between the pump field ($\omega_p$) and the cavity modes ($\omega_c$) and $U_0$ determines the interaction strength. Physically, $U_0$ represents the optical potential depth per photon in the cavity as well as the cavity mode frequency shift per particle. In general, $U_0$ can be complex but we will restrict our treatment to real $U_0$, meaning that we only consider dispersive atom-light interactions. 

We approximate the mode functions $\mathbf{f}_n$ in the interaction zone as plane waves, so that their polarization is constant. In the following we will only consider four different modes, hence we will change notation from $\{a_1,a_2,a_3,a_4\}\rightarrow\{\alpha_+,\alpha_-,\beta_+,\beta_-\}$ and $\{\mathbf{f}_1,\mathbf{f}_2,\mathbf{f}_3,\mathbf{f}_4\}\rightarrow\{\mathbf{f}_\alpha^+,\mathbf{f}_\alpha^-,\mathbf{f}_\beta^+,\mathbf{f}_\beta^-\}$, where
\begin{equation}
 \mathbf{f}^\pm_{\alpha,\beta}(\mathbf{x})=\exp({\pm\ii\mathbf{k}\!\cdot\!\mathbf{x}})\,\mathbf{e}_{\alpha,\beta}
 \label{eqn:modefun}
\end{equation}
and the polarization vectors fulfil the orthogonality relation $\mathbf{e}_\alpha\!\cdot\mathbf{e}_\beta=\delta_ {\alpha,\beta}$.
Two counterpropagating, orthogonally polarized modes $\mathbf{f}_\alpha^+$ and $\mathbf{f}_\beta^-$ are pumped with amplitudes $\eta_1\equiv\eta_+$ and $\eta_4\equiv\eta_-$, while the other two modes $\mathbf{f}_\alpha^-$ and $\mathbf{f}_\beta^+$ are only populated by scattered photons. This configuration represents only a slight change as compared to standard ring cavity cooling scheme~\cite{gangl2000cold,maes2007self,klinner2006normal}, but constitutes a very different situation physically. As the two counterpropagating pump fields do not interfere, no  prescribed optical lattice is formed and the system is inherently translation invariant. Note that imperfect mirrors in principle could lead to scattering between the two polarizations. Fortunately in a three mirror ring cavity the two orthogonal polarization modes are sufficiently frequency shifted due to the polarization dependent mirror reflection, so that no resonant scattering between the modes will occur. 

The force on a particle within the cavity field is given by the gradient of the optical dipole potential $\phi(x)=\hbar U_0|\mathbf{E}(\mathbf{x})|^2$ associated with the local field intensity, hence $m\ddot{\mathbf{x}_j}=-\nabla\phi(\mathbf{x}_j)$ with particle mass $m$. We restrict our treatment to the one dimensional motion along the cavity axis, so that $\mathbf{x}_j$ is replaced by $x_j$. Under these assumptions eqs.~\eqref{eqn:modesgeneral} and~\eqref{eqn:modefun} lead to:
\begin{subequations}
\begin{align}
\dot{\alpha}_+&=(\ii\delta-\kappa)\alpha_+
-\ii NU_0\,\theta\,\alpha_-+\eta_+ \label{eqn:mode1}\\
\dot{\alpha}_-&=(\ii\delta-\kappa)\alpha_-
-\ii NU_0\,\theta^*\,\alpha_+\label{eqn:mode2}\\
\dot{\beta}_+&=(\ii\delta-\kappa)\beta_+
-\ii NU_0\,\theta\,\beta_-\label{eqn:mode3}\\
\dot{\beta}_-&=(\ii\delta-\kappa)\beta_-
-\ii NU_0\,\theta^*\,\beta_++\eta_-\label{eqn:mode4},
\end{align}
\label{eqn:modes}
\end{subequations}
where $\theta=1/N\sum_ne^{-2\ii kx_j}$ defines the orderparameter and $\delta:=\Delta_c-N U_0$ is the effective cavity detuning .

These eqs.~\eqref{eqn:modes} describe two independent CARL geometries with different propagation directions~(\cf~\fref{fig:scheme}) which interact via the atomic density inhomogeneities.

The light induced optical potential explicitly reads:
\begin{multline}
\phi=\hbar U_0(\alpha_+\alpha_-^*\ee^{2\ii kx}+\alpha_+^*\alpha_-\ee^{-2\ii kx}\\
+\beta_+\beta_-^*\ee^{2\ii kx}+\beta_+^*\beta_-\ee^{-2\ii kx}).
\label{eqn:potential}
\end{multline}

For very large particle numbers the numerical simulation of equations of motion can be achieved only at large computational cost. However, in the limit $N\rightarrow\infty$ the dynamics of the gas can, for sufficiently short times be reliably approximated by a Vlasov equation (in 1D)~\cite{griesser2010vlasov, tesio2014kinetic}
\begin{equation}
\frac{\partial f}{\partial t}+v\frac{\partial f}{\partial x}-\frac{1}{m}\frac{\partial\phi}{\partial x}\frac{\partial f}{\partial v}=0
\label{eqn:vlasov}
\end{equation}
for the corresponding one-body phase space distribution function $f(x,v,t)$. Such a treatment misses, however, correlations in the density and field fluctuations which lead to cooling and heating on longer time scales~\cite{niedenzu2011kinetic}.
 Assuming periodic boundary conditions allows to restrict our treatment to the truncated phase space with $x\in(0,\lambda)$.
 
\section{Stability analysis}
Let us now investigate the coupled dynamics of the field modes and the gas described by eqs.~\eqref{eqn:modes}~--~\eqref{eqn:vlasov}. For a spatially homogeneous particle distribution the system is fully translation invariant and thus $\theta = 0$. For this reason the state defined by, $\alpha_-^0=\beta_+^0=0$ as well as
\begin{subequations}
\begin{gather}
f(x,v,t)=\lambda^{-1}F(v),\\
\alpha_+^0=\frac{\eta_+}{\kappa-i\delta}, \quad
\beta_-^0=\frac{\eta_-}{\kappa-i\delta}.
\end{gather}
\label{eqn:stationary}
\end{subequations}
constitutes a stationary solution of the system~\eqref{eqn:modes}-\eqref{eqn:vlasov}, regardless of the velocity distribution $F(v)$. Notice, in the following we only consider thermal (\ie Maxwell-Boltzmann) velocity distributions
\begin{equation}
F(v)=\frac{1}{\sqrt{\pi}v_\mathrm{T}}e^{-\left(\frac{v}{v_T}\right)^2}.
\label{eqn:thermdistr}
\end{equation}
Here we introduced the thermal velocity $v_\mathrm{T}$ which is connected to the temperature via $mv_\mathrm{T}^2/2=k_\mathrm{B}T$.

In this stationary state only forward scattering occurs without photon redistribution between the modes and thus there are no forces on the particles. Only deviations from perfect spatial homogeneity can lead to backscattering and the build-up of an optical lattice. To find out under which conditions such deviations are amplified and a subsequent phase transition to an ordered phase can occur,  we perform a linear stability analysis following Landau~\cite{landau1946on}. As a result we find that the steady state~\eqref{eqn:stationary} is unstable if and only if the dispersion relation $D(s)$ has at least one zero with a positive real part, where
\begin{equation}
D(s):=\delta^2+(s+\kappa)^2+\left[(s+\kappa) A-\ii\delta S\right]I(s).
\label{eqn:dispersion}
\end{equation}
In~\eqref{eqn:dispersion} we defined the total pump parameter $S$ and the pump asymmetry $A$ according to
\begin{equation}
S:=\abs{\eta_+}^2+\abs{\eta_-}^2 , \; ~~ A:=\abs{\eta_+}^2-\abs{\eta_-}^2.
\label{eqn:asym}
\end{equation}
Furthermore,
\begin{equation}
I(s):=\frac{N U_0^2v_R}{\kappa^2+\delta^2}\int_{-\infty }^{\infty } \frac{F'(v)}{s+2 \ii k v} \, \mathrm{d}v
\label{eqn:integral}
\end{equation}
with the recoil velocity~$v_\mathrm{R}=2\hbar k/m$. One finds that for every given pump asymmetry there exists a critical total pump parameter $S_c$ such that the homogeneous state is unstable for $S>S_c$ and stable otherwise.

For equal pump intensities, \ie $A=0$ , we recognize that~\eqref{eqn:dispersion} is almost exactly the same dispersion relation as one obtains for a transversally pumped ring-cavity. The only difference is that the wavenumber is multiplied by a factor 2 and the transversal pump-intensity is replaced by the sum of the two pump intensities $S$~\cite{griesser2010vlasov}. Obviously there exists a close analogy between the present setup with equal pump strengths and a transversally pumped ring-cavity. As in the latter case there appear stable selforganized solutions beyond an instability threshold~\cite{griesser2010vlasov}, which is given by
\begin{equation}
S_c^{A=0}:=\frac{k_BT}{NU_0^2}\frac{(\kappa^2+\delta^2)^2}{\hbar|\delta|}.
\label{eqn:thresdelta0}
\end{equation}
The other extreme case, $A=\pm S$, corresponds to a pure CARL instability~\cite{griesser2010vlasov} in which case no selforganization can take place.

The dependence of the critical pump parameter on the pump asymmetry can not be found in closed form. Nonetheless, solving $D(0+\ii\omega)=0$ for $A(\omega)$ and $S(\omega)$ yields the stability boundary using $\omega\in\mathbb{R}$ as parameter,~\cf~\fref{fig:phasediagr}.

The central question in the following is whether and under which conditions there occurs selforganization for nonzero pump asymmetries.

\section{Numerical simulation}
To gain deeper insight into the systems behaviour we have numerically solved the Maxwell-Vlasov equations~\eqref{eqn:modes}-\eqref{eqn:vlasov} for an initial condition close to homogeneous in space and a negative effective detuning $\delta$. These simulations confirm our predictions for the $A=0$ case. Furthermore they reveal that for sufficiently small $A$ and above threshold the system does evolve into a selfordered state, albeit one in which the gas possesses a non vanishing  centre of mass velocity $v_\mathrm{ph}$ constant in time. In the process of forming such a travelling wave the continuous translation symmetry is broken. As a matter of fact, the gas moves in the direction of the stronger pump beam.
\begin{figure*}[t]
\centering
\large a)\!\includegraphics[width=52mm]{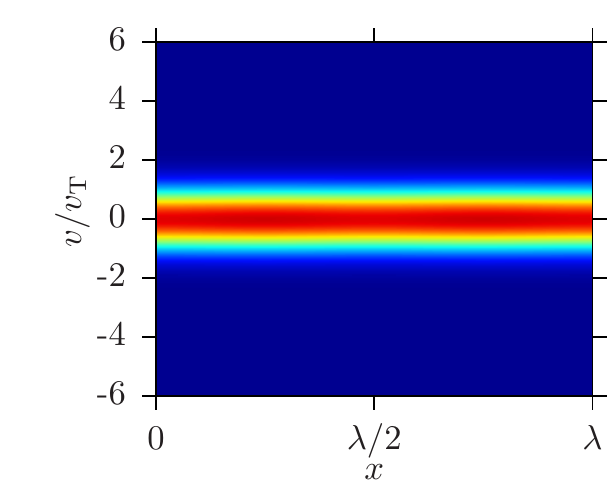}
\, ~~  b)\!\includegraphics[width=52mm]{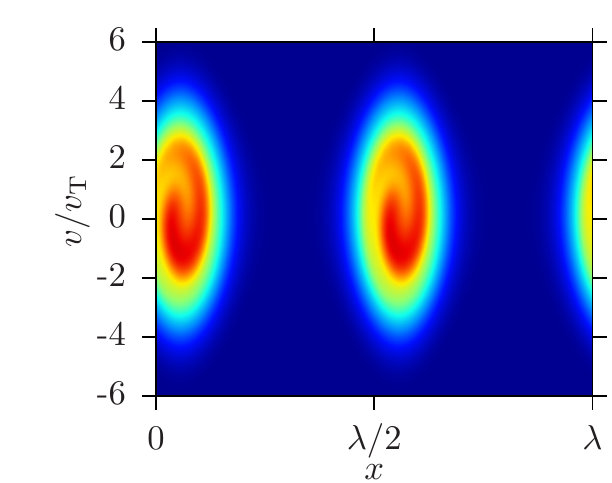}
\, ~~  c)\!\includegraphics[width=52mm]{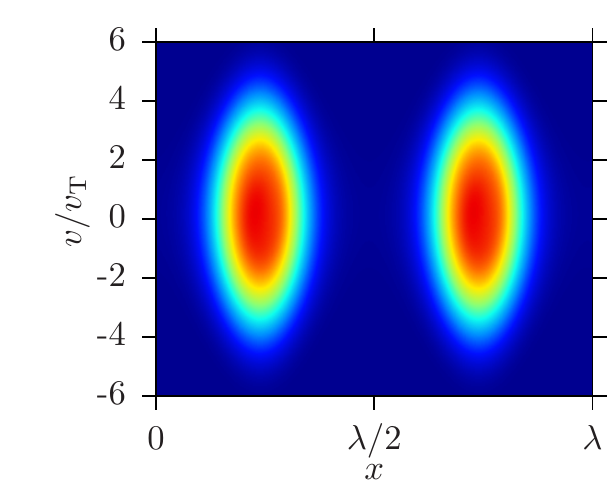}
\caption{(Colour online) Snapshots from the time evolution of the phase space density for a relative pump asymmetry $A/S=0.3$. The parameters are chosen to be: $k v_\mathrm{T}=1.5\kappa$, $N=2\!\cdot\!10^5$, the effective cavity detuning is set to $\delta=-\kappa$  and $U_0=-1/N$. a) Spatially homogeneous distribution at $t=0\kappa^{-1}$. b) During the ordering process at $t=15\kappa^{-1}$. c) Selfordered state at $t=32\kappa^{-1}$. Obviously, the system exhibits a centre of mass velocity but a spatially periodic order is formed.}
\label{fig:phsptevo1}
\end{figure*}
However, we find that, for a given $S$, as soon as $A$ exceeds a certain value, there still occurs a CARL instability resulting in a runaway centre of mass velocity. For an illustration of these processes,~\cf figs.~\ref{fig:phsptevo1}-~\ref{fig:vcm}.
\begin{figure*}[t]
\centering
\large a)\!\includegraphics[width=52mm]{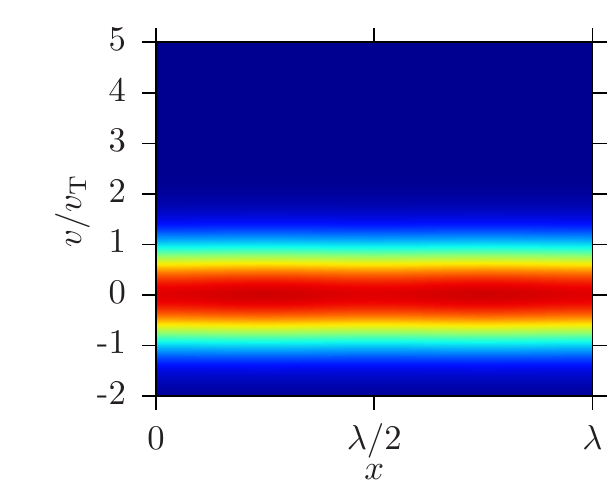}
\, ~~  b)\!\includegraphics[width=52mm]{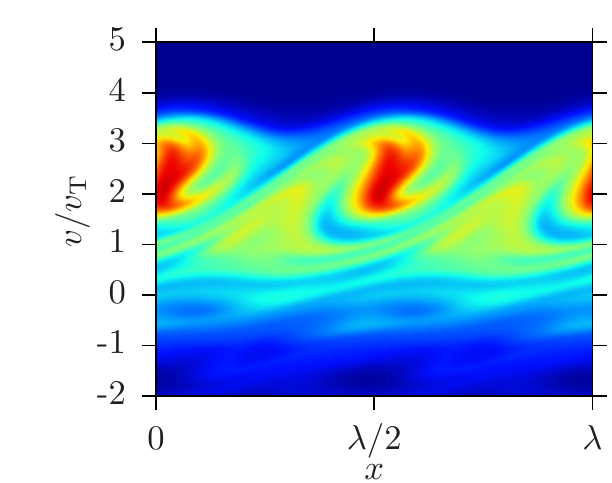}
\, ~~  c)\!\includegraphics[width=52mm]{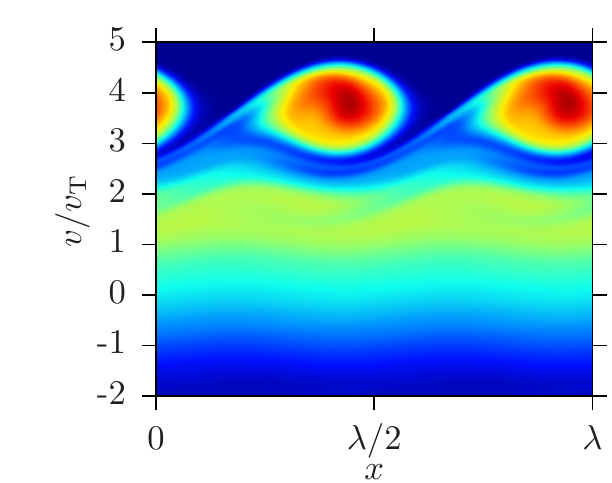}
\caption{(Colour online) Snapshots from the time evolution of the phase space density for a relative pump asymmetry $A/S=0.8$. The rest of the parameters are chosen to be equal to the ones of~\fref{fig:phsptevo1}. a) Spatially homogeneous distribution at $t=0\kappa^{-1}$. b) CARL instability at $t=10\kappa^{-1}$ and c) CARL instability at $t=20\kappa^{-1}$.}
\label{fig:phsptevo2}
\end{figure*}

While we have started from a particle ensemble at rest up to now and found a moving gas in a steady state, one can turn the idea around and use this setup to efficiently slowing down a cold atomic or molecular beam by collective scattering, improving a similar approach which has already been presented in~\cite{maes2007self} (see the red curve in~\fref{fig:vcm}).
\begin{figure}
\centering
\includegraphics[width=80 mm]{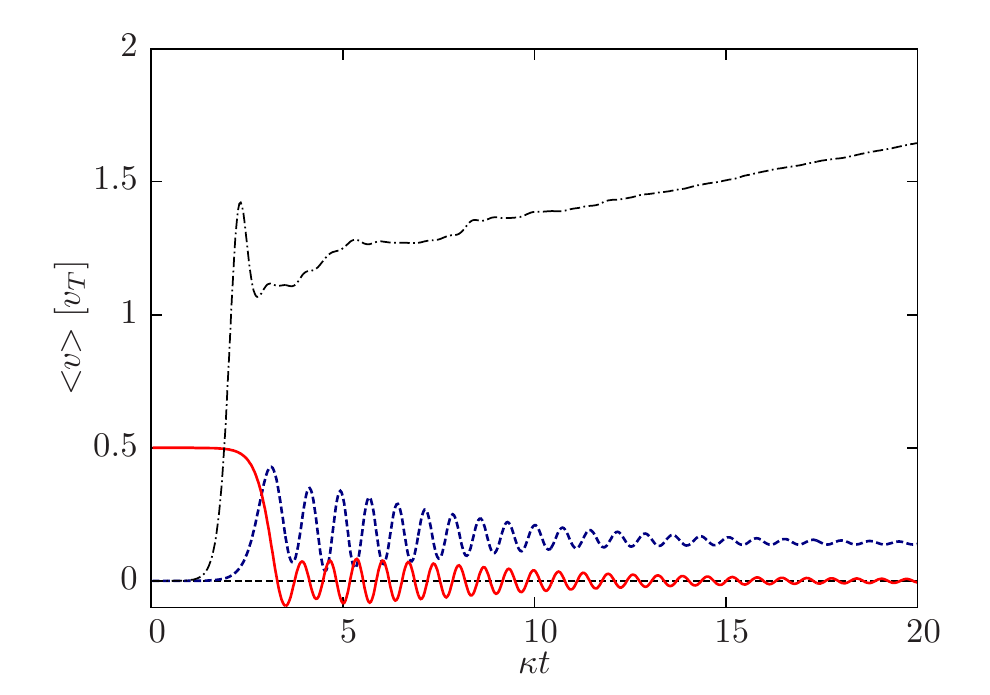}
\caption{(Color online) Evolution of the centre of mass velocity for the same parameters as in~\fref{fig:phsptevo1}. Sufficiently small $A/S=0.3$ (dashed blue) leads to a selforganized solution with a constant centre of mass velocity whereas for large $A/S=0.8$ (dash dotted black) the CARL phase is realised which results in an indefinitely increasing centre of mass velocity. The solid red curve shows the evolution of the mean velocity for a initial distribution with initially $<\!\!v\!\!>\neq0$ and $A=0$.}
\label{fig:vcm}
\end{figure}

\section{BGK-Waves}
As we have seen above, in the case of instability and depending on the pump asymmetry, the gas either enters the CARL regime, in which the centre of mass is accelerated indefinitely, or it settles in a selfordered, travelling-wave state with a constant phase velocity (\ie centre of mass velocity). Let us therefore investigate this latter type of solution more closely. From equation \eqref{eqn:vlasov} one deduces that any nonlinear wave with phase velocity $v_{\mathrm{ph}}$ must be of the BGK (Bernstein-Greene-Kruskal) form~\cite{bernstein1957exact}
\begin{equation}
f(x,v,t)=G\left(\frac{m (v-v_{\mathrm{ph}})^2}{2}+\phi(x,t)\right),
\label{eqn:BGK}
\end{equation}
where $G(.)$ is an arbitrary function. Furthermore, $\phi(x,t)$ may depend on $(x,t)$ only through $x-v_{\mathrm{ph}}t$, which implies that $\alpha_- e^{-2\ii kv_{\mathrm{ph}}t},\beta_+e^{2 \ii kv_{\mathrm{ph}}t}$ as well as $\alpha_+,\beta_-$ all be independent of time. To actually find the phase velocity from the equations of motion we require $G(.)$, which is obtained as the solution of an initial value problem and thus in general out of reach. Nevertheless it is possible to deduce a relationship between the phase velocity, the order parameter and the relative pump asymmetry in the form ($\Theta:=N|\theta|$)
\begin{equation}
\frac{A}{S}=\frac{-4\delta (\kappa^2+\delta^2-U_0^2\Theta^2) kv_{\mathrm{ph}}}{4(\kappa^2+\delta^2)k^2v_{\mathrm{ph}}^2+(\kappa^2+\delta^2-U_0^2\Theta^2)^2+(2\kappa U_0\Theta)^2}
\label{eqn:AoS}
\end{equation}
fulfilled by any nonlinear wave solution.
From~\eqref{eqn:AoS} we find that for $\delta<0$ the wave travels in the direction of the stronger pump beam, as long as $N|U_0|\leq\sqrt{\kappa^2+\delta^2}$. As soon as the inequality is violated, waves with sufficiently large order parameters propagate in the opposite direction. As such waves have never been observed numerically we have reason to expect them to be dynamically unstable. Hence we stipulate that the order parameter satisfies the bound $N|U_0||\theta|\leq \sqrt{\kappa^2+\delta^2}$.
\begin{figure}
\centering
\includegraphics[width=65mm]{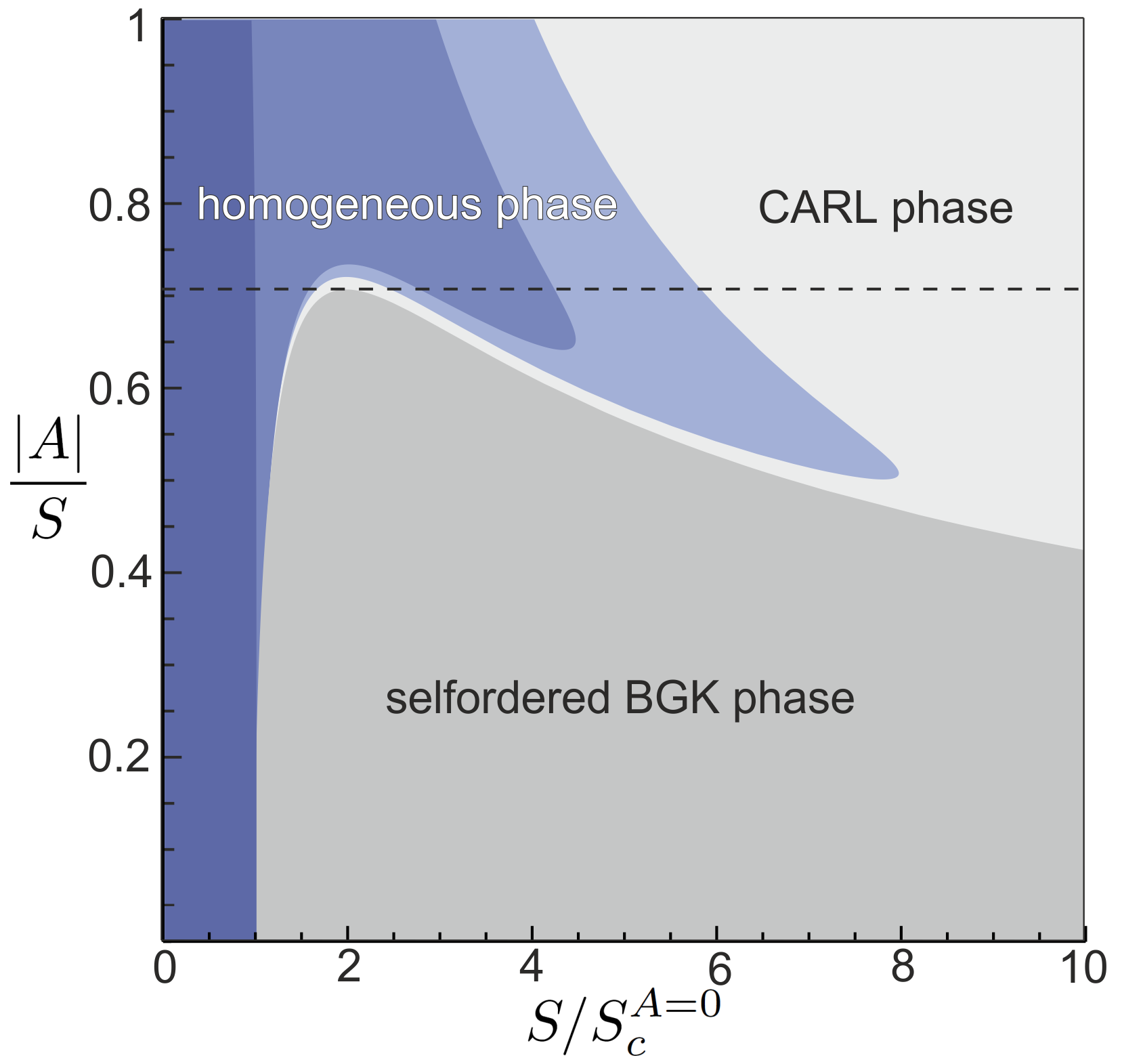}
\caption{Phase diagramm. The blue regions correspond to a stable homogeneous gas for different temperatures. From dark to light blue we have chosen $k v_T=1.5,\,50,\,100\,\kappa$. The dark grey region, bounded by~\eqref{eqn:BGKbound}, marks the parameter regime where a warm gas transitions to a BGK state~\eqref{eqn:BGK}. The dashed black line corresponds to the bound~\eqref{eqn:CARLbound}.}
\label{fig:phasediagr}
\end{figure}
Furthermore, eqn.~\eqref{eqn:AoS} allows to conclude that if
\begin{equation}
\frac{|A|}{S}>\frac{|\delta|}{\sqrt{\kappa^2+\delta^2}}
\label{eqn:CARLbound}
\end{equation}
there exists \emph{no} BGK-wave solution at all. This implies that if the homogeneous solution is unstable and the asymmetry exceeds the bound~\eqref{eqn:CARLbound}, the gas will \emph{necessarily} enter the CARL regime.

Eqn.~\eqref{eqn:AoS} can also be viewed as determining the necessary relative pump asymmetry $A/S$, which is needed to generate a wave with a prescribed phase velocity $v_\mathrm{ph}$ and order $|\theta|$. Notice, however, that the necessary total pump strength $S$ can not be inferred. In particular, in order to stop a beam (\ie to achieve $v_\mathrm{ph}=0$) the pump asymmetry has to be equal to zero.

The foregoing statements exhaust the characterization of the BGK solutions in absence of knowing $G(.)$.
Without going into details we state that for gas with temperature $kv_T\gg\kappa$, one finds that a BGK-wave will develop as soon as $S>S_{\mathrm{BGK}}$, where
 \begin{equation}
S_{\mathrm{BGK}}:=S_c^{A=0}\left[1+\left(\frac{\hbar A}{2k_BT}\frac{NU_0^2}{\sqrt{\kappa^2+\delta^2}}\right)^2\right].
\label{eqn:BGKbound}
\end{equation}
The results of the stability analysis and the discussion above are summarized in~\fref{fig:phasediagr}.

\section{Conclusions and outlook}
We demonstrated that utilizing orthogonally polarized counter propagating modes the physics of light induced selfordering is observable, similarly to the case of a transversely pumped ring resonator. In the case of no pump asymmetry the two setups are fully equivalent~\cite{griesser2010vlasov}. However, the system considered in this work is more versatile, because, in principle, by the choice of the pump asymmetry, ordered particle distributions with any prescribed centre of mass velocity can be generated. Therefore the control of the pump intensity allows for controlling the motion of gas particles inside the ring cavity. As a consequence, a particle beam can be effectively slowed down and trapped. Note that a different loss rate for the two polarization modes, as it often appears in practice, can be easily compensated by a correspondingly enlarged pump. 

Analogous physics should be present at zero temperature allowing to control and study degenerate quantum gases. Interesting effects can also be expected in the case of particles in optical lattices. Here collective scattering from orthogonally polarized modes can be used to gain insight into the particle quantum statistics at minimal perturbation or to induce tailored long-range interactions.

\begin{acknowledgments}
We thank C. Zimmermann for helpful discussions on experimental implementability of the system and acknowledge support by the Austrian Science Fund FWF through projects SFB FoQuS P13 and I1697-N27.
\end{acknowledgments}

%

\end{document}